# Evaluating the Efficiency of Software-only Techniques to Detect SEU and SET in Microprocessors

José Rodrigo Azambuja, Fernando Sousa, Lucas Rosa, Fernanda Lima Kastensmidt, *Member, IEEE*
Universidade Federal do Rio Grande do Sul (UFRGS) - Instituto de Informática
Av. Bento Gonçalves 9500, Porto Alegre - RS - Brazil
{jrfazambuja, faacsousa, lucas.rosa, fglima} @ inf.ufrgs.br

*Abstract*— This paper presents a detailed evaluation of the efficiency of software-only techniques to mitigate SEU and SET in microprocessors. A set of well-known rules is presented and implemented automatically to transform an unprotected program into a hardened one. SEU and SET are injected in all sensitive areas of a MIPS-based microprocessor architecture. The efficiency of each rule and a combination of them are tested. Experimental results show the inefficiency of the control-flow techniques in detecting the majority of SEU and SET faults. Three effects of the non-detected faults are explained. The conclusions can lead designers in developing more efficient techniques to detect these types of faults.

## I. INTRODUCTION

THE last-decade advances in the semiconductor industry have increased microprocessor performance exponentially. Most of this performance gain is due to smaller dimensions and low voltage transistors. However, the same technology that made possible all this progress also lowered the transistor reliability by reducing threshold voltage and tightening the noise margins [1, 2] and thus making them more susceptible to faults caused by energized particles [3]. As a consequence, high reliable applications demand fault-tolerant techniques capable of recovering the system from a fault with minimum implementation and performance overhead.

One of the major concerns is known as *soft error*, which is defined as a transient effect fault provoked by the interaction of energized particles with the PN junction in the silicon. This upset temporally charges or discharges nodes of the circuit, generating transient voltage pulses that can be interpreted as internal signals, thus provoking an erroneous result [4]. The most typical errors concerning soft errors are *single event upsets* (SEU), which are bit-flips in the sequential logic and *single event transients* (SET), which are transient voltage pulses in the combinatorial logic that can be registered by the sequential logic.

In areas where computer-based dependable systems are being introduced, the cost and development time are often major concerns. In such areas, highly efficient systems called systems-on-chip (SoC) are being used. SoC's are often designed using intellectual property (IP) cores and commercial off-the-shelf (COTS) microprocessors, which are only guaranteed to function correctly in normal environmental characteristics, while their behavior in the presence of soft errors is not guaranteed. Therefore, it is up to designers to harden their systems against soft errors. Fault tolerance by means of software techniques has been receiving a lot of attention on those systems, because they do not need any customization of the hardware.

Software implemented hardware fault tolerance (SIHFT) techniques exploit information, instruction and time redundancy to detect and even correct errors during the program flow. All these techniques use additional instructions in the code area to either recompute instructions or store and check suitable information in hardware structures. In the past years, tools have been implemented to automatically inject such instructions into C or assembly code, reducing significantly the costs.

Nevertheless, the drawbacks of software only techniques are the impossibility to achieve complete fault coverage [5], usual high overhead in memory and degradation in performance. Memory increases due to the additional instructions and often memory duplication, while the performance degradation comes from the execution of redundant instruction [6, 7, 8].

In this paper, the authors implemented a set of software-only techniques to harden a matrix multiplication algorithm in order to point out the main vulnerable areas that are not mitigated by these techniques, more specifically the ones affecting the control-flow. Results can guide designers to improve efficiency and detection rates of soft errors mitigation techniques based on software.

The paper is organized as follows: Section 2 presents related works on the area of software only techniques. Section 3 presents the software rules implemented and a tool that implements the automatic injection of such rules. Section 4 presents the fault injection campaign and results. Section 5 concludes the paper and presents future work.

## II. THE PROPOSED CASE-STUDY HARDENED PROGRAM METHODOLOGY

A set of transformation rules has been proposed in the literature. In [9], eight rules are proposed, divided in two groups: (1) aiming data-flow errors, such as data instruction replication [9, 10] and (2) aiming control-flow errors, such as Structural Integrity Checking [11], Control-Flow Checking by Software Signatures (CFCSS) [12], Control Flow Checking using Assertions (CCA) [13] and Enhanced Control Flow

Checking using Assertions (ECCA) [14]. The proposed techniques could achieve a full data-flow tolerance, concerning SEU's, being able to detect every fault affecting the data memory, which would lead the system to a wrong result. On the other hand, the control-flow techniques have not yet achieved full fault tolerance.

Most control-flow techniques divide the program into basic blocks by starting them in jump destination addresses and memory positions after branch instructions. The end of a basic block is on every jump instruction address and on the last instruction of the code.

ECCA extends CCA and is capable of detecting all the inter-BB control flow errors, but is neither able to detect intra-BB errors, nor faults that cause incorrect decision on a conditional branch. CFCSS is not able to detect errors if multiple BBs share the same BB destination address. In [15], several code transformation rules are presented, from variable and operation duplication to consistency checks.

Transformation rules have been proposed in the literature aiming to detect both data and control-flow errors. In [9], eight rules are proposed, while [16] used thirteen rules to harden a program. In this paper, we address six rules, divided into faults affecting the datapath and the controlpath.

### A. Errors in the Datapath

This group of rules aims at detecting the faults affecting the data, which comprises the whole path between memory elements, for example, the path between a variable stored in the memory, through the ALU, to the register bank. Every fault affecting these paths, as faults affecting the register bank or the memory should be protected with the following rules:
- Rule #1: every variable used in the program must be duplicated;
- Rule #2: every write operation performed on a variable must be performed on its replica;
- Rule #3: before each read on a variable, its value and its replica's value must be checked for consistency.

Figure 1 illustrates the application of these rules to a program with 3 instructions. Instructions 1, 3, 7 and 8 are inserted due to rule #3, while instruction 3, 6 and 10 are inserted due to rules #1 and #2.

| | |
|---|---|
| ld r1, [r4] | 1: bne r4, r4', error<br>2: ld r1, [r4]<br>3: ld r1, [r4 + offset] |
| add r1, r2, 1 | 4: bne r2, r2', error<br>5: add r1, r3, 1<br>6: add r1', r3, 1 |
| st [r1], r2 | 7: bne r1, r1', error<br>8: bne r2, r2', error<br>9: st [r1], r2<br>10: st [r1 + offset], r2 |

Figure 1: datapath rules

Combined, these techniques duplicate the used data size, such as number of registers and memory addresses and, therefore, the microprocessor must have spare registers and the memory must have spare memory positions. This issue can also be solved by setting the compiler options to restrict the program to a given number of registers and restrict the data section.

### B. Errors in the Controlpath

This second group of rules aims at protecting the program's flow. Faults affecting the controlpath usually cause erroneous jumps, such as an incorrect jump address or a bitflip in a non-jump instruction's opcode which becomes a jump instruction. To detect these errors, three rules are used in this paper:
- Rule #4: every branch instruction is replicated on both destination addresses.
- Rule #5: an unique identifier is associated to each basic block in the code;
- Rule #6: At the beginning of each basic block, a global variable is assigned with its unique identifier. On the end of the basic block, the unique identifier is checked with the global variable.

Branch instructions are more difficult to duplicate than non-branch instructions, since they have two possible paths, when the branch condition is true or false. When the condition is false, the branch can be simply replicated and added right after the original branch, because the injected instruction will be executed after the branch. When the condition is taken, the duplicated branch instruction must be inverted and inserted on the branch taken address.

Figure 2 illustrates rule #4 applied to a simple program. For the branch if equal (BEQ) instruction, instructions 2, 4 and 5 must be inserted to replicate it, where instruction 5 is the inverted branch (branch if not equal). Instruction 4 is necessary to avoid false error alerts.

| | |
|---|---|
| beq r1, r2, 6 | 1: beq r1, r2, 5<br>2: **beq r1,r2, error** |
| add r2, r3, 1 | 3: add r2, r3, 1 |
| | **4: jmp 6**<br>**5: bne r1,r2, error** |
| add r2, r3, 9<br>jmp end | 6: add r2, r3, 9<br>7: jmp end |

Figure 2: rule #4

| | |
|---|---|
| beq r1, r2, 6 | 1: beq r1, r2, 6 |
| add r2, r3, 1 | **2: mv rX, signature 1**<br>3: add r2, r3, 1<br>**4: bne rX, signature 1, error** |
| add r2, r3, 9<br>st [r1], r2 | **5: mv rX, signature 2**<br>6: add r2, r3, 9<br>7: st [r1], r2<br>**8: bne rX, signature 2, error** |
| jmp end | 9: jmp end |

Figure 3: rules #5 and #6

The role of rules #5 and #6 is to detect every erroneous jump in the code. They achieve this by inserting a unique identifier to the beginning of each basic block and checking its value on its end. Figure 3 illustrates a program divided in two basic blocks (instructions 2-4 and 5-8). Instructions 2 and 5 are inserted to set the signature, while instructions 4 and 8 are inserted to compare the signatures with the global value.

### III. FAULT INJECTION EXPERIMENTAL RESULTS

The chosen case-study microprocessor is a five-stage pipeline microprocessor based on the MIPS architecture, but with a reduced instruction set. The miniMIPS microprocessor is described in [17]. In order to evaluate both the effectiveness and the feasibility of the presented approaches, an application based on 6x6 matrix multiplication algorithm is used.

A tool called PosCompiler was developed to automate the software transformation. The tool receives as input the program's binary code and therefore is compiler and language independent and is capable of implementing the presented rules, divided in 3 groups. The first group, called variables, implements rules #1, #2 and #3; group 2, called inverted branches, implements rule #4 and, finally, group 3, also known as signatures, implements rules #5 and #6. The user is allowed to combine the techniques in a graphical interface.

We generated through PosCompiler four hardened programs, implementing: (1) signatures, (2) variables, (3) inverted branches and (4) signatures, variables and inverted branches combined. Table 1 shows the original and modified program's execution time, cod size and data size.

Table 1: original and hardened program's characteristics

| Source | Original | (1) | (2) | (3) | (4) |
|---|---|---|---|---|---|
| Exec. Time (ms) | 1.24 | 1.40 | 2.55 | 1.30 | 2.71 |
| Code Size (byte) | 2060 | 3500 | 4340 | 2580 | 6012 |
| Data Size (byte) | 524 | 532 | 1048 | 524 | 1056 |

First, thousands of faults were injected in the non-protected microprocessor, one by program execution. At the end of each execution, the results stored in memory were compared with the expected correct values. If the results matched, the fault was discarded. The amount of faults masked by the program is application related and it should not interfere with the analysis.

When 100% signal coverage was achieved and at least 3 faults per signal were detected we normalized the faults, varying from 3 to 5 faults per signal, and those faults build the test case list.

In order to achieve a detailed fault analysis, we sorted the faults by their source and effect on the system. We defined four groups of fault sources to inject SEU and SET types of faults: datapath, controlpath, register bank and ALU. We assumed the program and data memories are protected by Error Detection and Correction (EDAC) and therefore faults in the memories were not injected.

The fault effects were classified into 2 different groups: program data and program flow, according to the fault effect. To sort the faults among these groups, we continuously compared the Program Counter (PC) of a golden microprocessor with the PC of the faulty microprocessor. In case of a mismatch, the injected fault was classified as flow effect. If the PC matched with the golden's, the fault was classified as a data effect.

When transforming the program, new instructions were added and therefore the time in which the faults were injected changed. Since the injection time is not proportional to the total execution time, we mapped each fault locating the instruction where the fault was injected (by locating its new PC) and pipeline stage where the fault was manifested. Around 1% of the total number of faults could not be mapped and were changed by new faults.

Results show that the technique called variables (2) presented the highest detection rate among the three. It was capable of detecting all the faults injected in the register bank and the faults that caused errors on the data flow. The ALU was not completely protected because it also has control flow signals and some of these signals affected the program's flow. With 110% code size overhead, this technique could detect 77% overall. Technique (3) was able to complement technique (1) by detecting faults on branch instructions, mainly in the ALU, where most of the branch errors were found. By using technique (2) and (3), with 135% in code size overhead, these techniques combined could detect 79% overall.

The signatures (1), on the other hand, were responsible for detecting the faults affecting the program's flow, but it could not reach a high detection rate.

When combining all of them, the fault detection coverage reaches 80% with a code size increase of 192% and execution time increase of 118%. However, 20% of faults remain undetected.

Table 3: (1) signatures, (2) variables, (3) inverted branches and (4) signatures, variables and inverted branches combined.

| | Source | # of Signals | Data | Hardened program versions (%) | | | | Flow | Hardened program versions (%) | | | |
|---|---|---|---|---|---|---|---|---|---|---|---|---|
| | | | | (1) | (2) | (3) | (4) | | (1) | (2) | (3) | (4) |
| SET | Reg. Bank | 2 | 9 | - | 100 | - | 100 | 1 | - | 100 | - | 100 |
| | ALU | 10 | 22 | - | 100 | - | 100 | 21 | - | 52.3 | 28.5 | 80.8 |
| | Controlpath | 29 | 90 | - | 100 | - | 100 | 46 | 2.17 | 23.9 | 2.17 | 23.9 |
| | Datapath | 8 | 37 | - | 100 | - | 100 | 3 | - | 100 | - | 100 |
| | Total | 49 | 158 | - | 100 | - | 100 | 71 | 1.4 | 36.7 | 9.8 | 45.1 |
| SEU | Reg. Bank | 36 | 18 | - | 100 | - | 100 | 18 | - | 100 | 5.5 | 100 |
| | ALU | 2 | 2 | - | 100 | - | 100 | 0 | - | - | - | - |
| | Controlpath | 126 | 63 | - | 100 | - | 100 | 56 | 1.8 | 17.8 | 1.7 | 19.6 |
| | Datapath | 20 | 19 | - | 100 | - | 100 | 1 | - | - | 100 | 100 |
| | Total | 184 | 102 | - | 100 | - | 100 | 75 | 1.3 | 37.3 | 4 | 40 |

## IV. ANALYZING THE UNDETECTED FAULTS

Technique (2) presented a high detection rate for data effects and faults injected in the datapath and register bank, while (3) protected the branch instructions and complemented (1). The techniques combined (4) also presented and interesting result, which is that these techniques can be scaled in order to achieve a higher detection rate. On the other hand, the technique (1) presented a detection rate below expected and this result will be analyzed in this session.

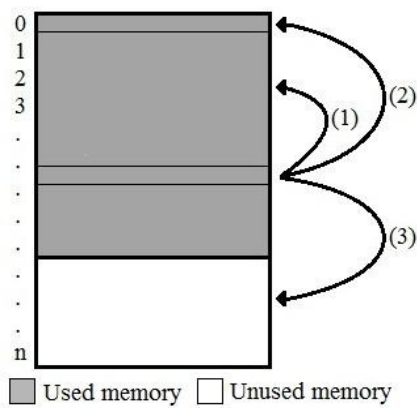

Figure 4: signatures drawbacks

The signatures have three major drawbacks, which caused the low detection rate. The first one is incorrect jumps to the same basic block (intra-block), which cannot be detected since the unique identifier is an invariant and therefore does not depend on the instructions. The application used in this paper spends 83% of its time on the same basic block, which occupies 20% of the program data, and therefore increases the occurrence of this drawback, that can be seen on figure 4, (1).

The second drawback is incorrect jumps to the beginning of a basic block that also cannot be detected, because the global variable containing the unique identifier is updated in the beginning of the basic blocks. The occurrence of such error is proportional to the number of basic blocks per instructions, which is higher in control-flow applications. Also, some microprocessors, such as the used in this paper, have a fault detection mechanism that resets themselves in some cases, such as when an inexistent instruction is fetched. This drawback can be seen on figure 4, (2).

Finally, the last drawback is incorrect jumps to unused memory positions, which are filled with NOP instructions. To correct this drawback, a watchdog with a timer could be used. This drawback can be seen on figure 4, (3).

Table 3: signatures drawbacks

| Drawback | Same Basic Block (1) | Beginning of Basic Block (2) | Unused Memory (3) |
|---|---|---|---|
| (%) | 37.4 | 42.2 | 20.4 |

Table 3 shows the distribution of undetected faults among the three drawbacks.

## V. CONCLUSIONS

In this paper we presented a set of rules based on software-only techniques to detect soft errors in microprocessors. A set of faults was built and a fault injection campaign was realized on the implemented techniques. Results showed that the variables and inverted branches presented a high detection rate, up to 77%, while the signatures showed results below expected. The signatures were then analyzed and three drawbacks were found explaining the undetected faults.

We are currently working on improving the detection rates and decreasing the impact of the drawback on the signatures technique.